\documentclass[aps,prd,twocolumn,superscriptaddress,showpacs,amsmath,amssymb,nofootinbib, preprintnumbers]{revtex4-1}
\usepackage{graphicx}
\usepackage{epstopdf}
\usepackage{float}
\usepackage{color}
\usepackage[toc,page]{appendix}
\usepackage[usenames,dvipsnames]{xcolor}

\newcommand{\be}{\begin{equation}}
\newcommand{\ee}{\end{equation}}
\newcommand{\ba}{\begin{eqnarray}}
\newcommand{\ea}{\end{eqnarray}}

\newcommand{\beq}{\begin{equation}}
\newcommand{\eeq}{\end{equation}}
\newcommand{\beqa}{\begin{eqnarray}}
\newcommand{\eeqa}{\end{eqnarray}}
\newcommand{\nn}{\nonumber}

\begin{document}
\title{Ultraspinning limits and rotating hyperboloid membranes}

\author{Robie A. Hennigar}
\email{rhennigar@uwaterloo.ca}
\affiliation{Department of Physics and Astronomy, University of Waterloo,
Waterloo, Ontario, Canada, N2L 3G1}

\author{David Kubiz\v n\'ak}
\email{dkubiznak@perimeterinstitute.ca}
\affiliation{Perimeter Institute, 31 Caroline St. N., Waterloo,
Ontario, N2L 2Y5, Canada}
\affiliation{Department of Physics and Astronomy, University of Waterloo,
Waterloo, Ontario, Canada, N2L 3G1}

\author{Robert B. Mann}
\email{rbmann@sciborg.uwaterloo.ca}
\affiliation{Department of Physics and Astronomy, University of Waterloo,
Waterloo, Ontario, Canada, N2L 3G1}

\author{Nathan Musoke}
\email{nmusoke@perimeterinstitute.ca}
\affiliation{Perimeter Institute, 31 Caroline St. N., Waterloo,
Ontario, N2L 2Y5, Canada}

\begin{abstract}
We apply the hyperboloid membrane limit  to the general Kerr-AdS metrics and their recently studied super-entropic cousins and obtain a new class of rotating black holes, for which the rotational parameters in multiple directions attain their maximal value---equal to the AdS radius.
These new solutions have a potential  application in the description of
holographic fluids with vorticity. They also possess interesting thermodynamic properties:  we show that---despite the absence of Misner strings---the Bekenstein--Hawking entropy/area law is still violated, raising a question about the origin of this violation.

\end{abstract}

\date{\today}
\maketitle

\section{Introduction}
The study of general relativity in higher dimensions has remained an active and popular field, in part motivated by its applicability in string theory and various gauge/gravity dualities, and also due to the deeper perspective it provides for our understanding of four dimensional gravity \cite{Emparan:2008eg}.  Of particular interest are higher dimensional black hole solutions, whose physics highlights the peculiarities of four dimensions. For example, unlike their four-dimensional cousins described by the Kerr solution, the higher dimensional asymptotically flat, rotating black holes of Myers and Perry solve  Einstein's equations with arbitrarily large angular momentum for a given mass \cite{Myers:1986un}.  When possessing large angular momentum, these black holes are said to be \textit{ultraspinning}.

Ultraspinning limits of rotating black holes have been the attention of numerous studies.  Emparan and Myers~\cite{Emparan:1995a} found that in $d \ge 6$, ultraspinning Myers--Perry black holes present an instability reminiscent of the Gregory--Laflamme instability for black branes \cite{Gregory:1993vy}.  Arguably more interesting is the case of Kerr black holes in asymptotically anti de Sitter (AdS) space, where the `ultraspinning limit' corresponds to the case where the rotation parameter, $a$, approaches the AdS radius, $l$.\footnote{In what follows we shall take this as a definition of the ultraspinning limit of rotating AdS black objects. Note, however, that this is different from the definition used in \cite{Armas:2010hz} and the resulting solutions do not necessarily possess (for a given mass) large angular momentum.}
Here a variety of ultraspinning solutions can be obtained depending on how the
\be
a \to l
\ee
limit is performed.   Caldarelli et al. \cite{Caldarelli:2008pz} studied the \textit{black brane limit} wherein the physical mass is held fixed and the $a\to l$ limit is taken while simultaneously zooming in to the pole.  This limit is sensible only in $d \ge 6$ and yields a static, asymptotically flat black brane.  Armas and Obers later demonstrated that the same limiting solution is obtained by taking $a \to \infty$ while keeping the ratio $a/l$ fixed, their approach having the advantage of being directly applicable to dS solutions as well \cite{Armas:2010hz}. Caldarelli et al. have also studied the \textit{hyperboloid membrane limit} where the horizon radius, $r_+$, is held fixed while zooming in to the pole and taking $a \to l$ \cite{Caldarelli:2008pz, Caldarelli:2012cm}.  This limit, sensible for  $d \ge 4$, yields a rotating AdS hyperboloid membrane with horizon topology $\mathbb{H}^2 \times S^{d-4}$.    Klemm et al. and the present authors have studied the \textit{super-entropic limit} in which one works in rotating coordinates, rescales an azimuthal coordinate to absorb any singular quantities from the $a \to l$ limit, and then compactifies the resulting azimuthal coordinate \cite{Gnecchi:2013mja, Klemm:2014rda, Hennigar:2014cfa, Hennigar:2015cja}.  This limit, sensible for $d \ge 4$, is so-named because
the resulting solutions are super-entropic for some values of the parameters, that is, their entropy exceeds, for a given thermodynamic volume, the bound set by the \textit{reverse isoperimetric inequality} \cite{CveticEtal:2010}.

Recently, ultraspinning AdS black holes have found application via the \textit{fluid/gravity correspondence}.  This powerful duality relates the gravitational degrees of freedom residing on the boundary of a $d$-dimensional asymptotically AdS spacetime to the hydrodynamics of a relativistic fluid in $d-1$ spacetime dimensions (see, e.g., review \cite{Hubeny:2011hd}).  On the hydrodynamics side, this paradigm has been used to understand the role of quantum anomalies in hydrodynamical transport \cite{Son:2009tf} as well as understanding the fluid dynamics of superfluids \cite{Herzog:2009xv}. On the gravity side it has been utilized in the construction of static, inhomogeneous black hole solutions \cite{Bhattacharyya:2008mz} and as a means to classify the possible black hole solutions of Einstein--Maxwell-AdS theory \cite{Klemm:2014nka}.

As progress continues to be made, there are indications that this formalism may prove useful as a means of understanding rotating Bose/Fermi gases via holography \cite{Fetter:2009zz} as well as turbulence and wave propagation in metamaterials \cite{Leonhardt:2000fd}.  Important advances in this direction have been spearheaded by Leigh et al. \cite{Leigh:2011au} and elaborated upon in later research \cite{Caldarelli:2012cm, Leigh:2012jv, Petropoulos:2014yaa}, where a holographic treatment of fluids with vorticity was given.  Within this context, it was found that the conformal boundary of the spacetime obtained via the hyperboloid membrane limit of the $4d$ Kerr-AdS solution is of Papapertrou--Randers form and and can be cast in the Zermelo frame \cite{Caldarelli:2012cm}.  Such metrics are referred to as acoustic/optical since they describe the propagation of sound/light in relativistic fluids \cite{Gibbons:2008zi}, and feature prominently in descriptions of analogue gravity \cite{Barcelo:2005fc}.

In this  paper
we apply the hyperboloid membrane limit to the general Kerr-AdS metrics in all dimensions, thereby obtaining new ultraspinning black hole solutions.  We find that for a black hole with $m$ rotational parameters it is possible to perform the hyperboloid membrane limit successively  $m$ times in even dimensions and $m-1$ times in odd dimensions.  We further show that the hyperboloid membrane limit can be combined with the super-entropic limit, allowing us to construct black holes with all  rotation parameters equal to the AdS radius in any dimension.
The obtained solutions feature interesting thermodynamics as well as, their boundary being a `generalization of a rotating Einstein universe',
may prove useful for the study of (higher-dimensional) holographic fluids with vorticity.

\section{Singly spinning hyperboloid membranes}\label{Sec2}

 We begin by briefly recapitulating the hyperboloid membrane solutions of \cite{Caldarelli:2008pz, Caldarelli:2012cm} and investigating their thermodynamics. Consider the singly-spinning Kerr-AdS solution in $d$ spacetime dimensions \cite{Hawking:1998kw},
\ba\label{KerrAdSsingle}
ds^2 &=& -\frac{\Delta_a}{\rho^2_a}\left[dt - \frac{a}{\Xi_{{a}}} \sin^2\!\theta d\phi \right]^2 + \frac{\rho^2_a}{\Delta_a}dr^2 + \frac{\rho^2_a}{\Sigma_a}d\theta^2\qquad  \\
&+& \frac{\Sigma_a \sin^2\!\theta }{\rho_{{a}}^2} \left[a dt - \frac{r^2+a^2}{\Xi_{{a}}} d\phi \right]^2 + r^2 \cos^2\!\theta d\Omega_{d-4}^2\,, \nn
\ea
where
\ba
\Delta_a &=& (r^2+a^2)(1+\frac{r^2}{l^2}) - 2mr^{5-d}, \quad \Sigma_a = 1-\frac{a^2}{l^2}\cos^2\!\theta\,,\nn\\
\Xi_{{a}} &=& 1-\frac{a^2}{l^2}, \quad \rho^2_a = r^2+a^2\cos^2\!\theta\,.
\ea
This metric is well defined only for $a^2 < l^2$.  However, one way to obtain a finite $a^2 = l^2$ limit is to perform the limit while keeping the horizon radius finite and zooming in to the pole via
\be\label{eq:hyperlim_subs}
\sin \theta = \sqrt{\Xi_a} \sinh \frac{\sigma}{2}
\ee
keeping the new coordinate $\sigma\in[0,\infty)$ constant.
The resultant metric is given by
\ba\label{onespinlimit}
	ds^2 =&& -f \left(dt - l\sinh^2(\sigma/2) d\phi \right)^2 + \frac{dr^2}{f}
	\nn\\
	&&+ \frac{\rho^2}{4} \left(d\sigma^2 + \sinh^2\!\sigma d\phi^2 \right) +r^2 d\Omega_{d-4}^2\,,
\ea
where
\ba
\rho^2 = r^2+l^2 \,,  \quad f = 1 +\frac{r^2}{l^2} - \frac{2 m}{\rho^2 r^{d-5}}\,.
\ea
The horizon is located at $r=r_+$, the largest real root of $f(r)$.  Here the horizon clearly has topology $\mathbb{H}^2 \times S^{d-4}$, and so these solutions represent rotating black hyperboloid membranes,
c.f. `{\em rotating topological black holes}' with horizon topology $\mathbb{H}^2 \times \mathbb{H}^{d-4}$  reviewed in App.~A.

 The metrics \eqref{onespinlimit}  are asymptotically  AdS.
To obtain the boundary metric for \eqref{onespinlimit}, we rescale by $(l/r)^2$ and take the limit $r \to \infty$ obtaining
\ba
ds^2_{\rm bdry} &=& -(dt - l \sinh^2(\sigma/2) d\phi)^2
\nn\\
&&+ \frac{l^2}{4}( d\sigma^2 + \sinh^2\!\sigma d\phi^2) + l^2 d\Omega^2_{d-4} \, .
\ea
The $(t, \sigma, \phi)$ sector (which is a timelike fibration over $\mathbb{H}^2$) has constant negative curvature satisfying
\be
R_{\mu\nu\rho\lambda} = -\frac{1}{l^2}\left(g_{\mu \rho}g_{
\nu\lambda} - g_{\mu \lambda}g_{\nu \rho} \right) \, .
\ee
In particular this means that in the $4d$ case, the boundary is nothing more than AdS$_3$,  and more generally, AdS$_3 \times S^{d-4}$ for $d > 4$. Due to its connection with the Kerr-AdS solution, this boundary can be thought of as an Einstein static universe which rotates effectively at the speed of light \cite{Hawking:1998kw}.

The  obtained solutions \eqref{onespinlimit} are a particular case of the `nutty' spacetimes constructed in \cite{Mann:2003zh, Mann:2005ra}.  To see this,  let us specify to four dimensions and consider the hyperbolic Taub-NUT-AdS metric \cite{Bardoux:2013swa},
\ba\label{taub-nut}
ds^2 &=& -f\left(dt - 4n \sinh^2(\theta/2)d\phi \right)^2 + \frac{d\rho^2}{f}
\nn\\
&&+ (\rho^2+n^2)(d\theta^2 + \sinh^2\!\theta d\phi^2)\,,
\nn\\
f &=& \frac{n^2-\rho^2 -2M\rho + l^{-2}(\rho^4 + 6n^2 \rho^2 - 3n^4 )}{\rho^2+n^2}\, .
\ea
Here there are no Misner strings, the fibration is trivial, and no Euclidean NUT solutions (with a zero-dimensional fixed point set) exist.
The metric \eqref{onespinlimit} in $d=4$ is simply a special case of the above hyperbolic Taub-NUT-AdS solution with
\be \label{rescaling}
 n=l/2\,,\quad \rho = r/2 \, , \quad M=m/8\, .
\ee
Similar identifications with (single NUT parameter) hyperbolic Taub-NUT-AdS metrics can be made in higher dimensions.

The solution \eqref{onespinlimit}  describes a  rotating black hyperboloid membrane,  whose horizon structure depends on the value of $m$.  In 4d if $m > 8 \sqrt{3} l /9$ there is an inner and outer horizon, when $m = 8 \sqrt{3} l /9$ the horizons coincide and the black hole is extremal, while for $m < 8 \sqrt{3} l /9$ there is no horizon.  In $5d$ there is a single horizon provided $m > l^2/2$.  In $d > 5$ there is a single horizon provided that $m > 0$.

 It is interesting to note that, while the Lorentzian spherical Taub-NUT solution suffers from closed timelike curves (CTCs), this is not a generic feature in the hyperbolic case.  For $n \le l/2$  any CTCs in the metric \eqref{taub-nut}  will be hidden behind an event horizon \cite{Bardoux:2013swa}.  This feature remains true for our metric \eqref{onespinlimit} for all $d \ge 4$.   The metric \eqref{onespinlimit} is free from curvature singularities in the cases $d=4,5$ but for $d \ge 6$ the Kretschmann scalar diverges at the origin.

The angular velocity of  \eqref{onespinlimit} at any $(r,\sigma)$ is given by
\be
\omega = -\frac{g_{t\phi}}{g_{\phi\phi}} = \frac{4l f \sinh^2(\sigma/2)}{4fl^2\sinh^4(\sigma/2) - (r^2+l^2)\sinh^2\!\sigma}
\ee
implying that the horizon ($f=0$) does not rotate, whereas the boundary rotates with $\omega=\Omega_\infty = -1/l$.
As we shall see,
the combination that should enter the first law of thermodynamics is $\Omega = \Omega_h-\Omega_\infty = 1/l$, as is well-known in other contexts \cite{Gibbons:2004ai}.

The thermodynamics of these solutions
are difficult to study for a number of reasons.  The applicability of the standard Euclidean action formalism for obtaining thermodynamic quantities is questionable, since here the Euclidean version of the metric \eqref{onespinlimit} is asymptotically de Sitter, rather than AdS.  Furthermore, the thermodynamic quantities derived from a Euclidean action necessarily satisfy the first law of thermodynamics, and so it would be desirable to reproduce these quantities via an independent means.  However, the integrals for conserved charges are divergent since the hyperboloid is non-compact.\footnote{ Note that it is not possible to compactify the $(\sigma, \phi$) sector due to rotation \cite{Klemm:1998kd}.}

Consequently the best one can do is to write  expressions for the divergent thermodynamic quantities in terms of thermodynamic densities.  In doing so, it is appropriate to transform to a coordinate system that does not rotate at infinity, which can be obtained by transforming $\phi \to \phi - t/l$.\footnote{ The existence of this   non-rotating frame is somewhat surprising:  the more general hyperbolic Taub-NUT-AdS solution (with $n\neq l/2$) does not rotate uniformly---its  rotation is $\theta$-dependent and concentrates in the central region of the brane \cite{Bardoux:2013swa}.}
Working in these non-rotating coordinates and using the method of conformal completion \cite{Ashtekar:1999jx}, we have found that the conserved charges associated with the $\partial_t$ and $\partial_\phi$ Killing vectors are
\ba\label{thermos}
M &=& \frac{ m \omega_{d-4} }{16}  \int \left[(d-1) \cosh^2 (\sigma/2) -1\right]\sinh \sigma d\sigma \, ,
\nn\\
J &=& \frac{(d-1) ml }{16} \omega_{d-4} \int \sinh^2 (\sigma/2) \sinh \sigma d\sigma \, ,
\nn\\
A &=& \frac{\pi (r_+^2+l^2)r_+^{d-4}}{2} \omega_{d-4} \int \sinh \sigma d\sigma \, ,
\ea
where  $\omega_{d-4}$ is the area of a $(d-4)$-dimensional unit sphere, the integrals are defined on $\sigma \in [0, \infty)$,  and we have included the horizon area, $A$. The same results are obtained using Komar integration with $m=0$ background subtraction.   In what follows we shall introduce a `cutoff' $\Upsilon$ and integrate over $\sigma\in [0,\Upsilon]$. With the identifications \eqref{S1} or \eqref{S2} (see below), the first law \eqref{firstlaw} is valid for any $\Upsilon$.

Obviously the hyperboloid membranes do rotate and, having a non-trivial $\Omega$, the corresponding $\Omega \delta J$ term has to appear in the first law. Surprisingly, such a term is often ignored when thermodynamics of their topological Taub-NUT-AdS cousins are considered, e.g. \cite{Astefanesei:2004ji, Astefanesei:2004kn, Lee:2015wua}.
A temperature can be obtained by calculating the surface gravity at the horizon,
\be
\kappa^2 = (\nabla_\mu L)(\nabla^\mu L)\,,
\ee
where $-L^2 = \xi_\mu \xi^\mu$ and $\xi = \partial_t + \Omega_h \partial_\phi$ is the null generator of the horizon.  Carrying out these computations (the result is of course the same in both the rotating and non-rotating coordinates), we find
\be
T = \frac{\kappa}{2 \pi} =\frac{f'}{4\pi} = \frac{(d-5)l^2 + (d-1)r_+^2}{4 \pi r_+ l^2} \, .
\ee

Curiously,  these quantities do not satisfy the first law of thermodynamics
\be\label{firstlaw}
\delta M=T\delta S+\Omega\delta J\,,
\ee
if the entropy $S$ is identified in the standard way as one-quarter the horizon area.  If instead we force the first law to hold with the above quantities and determine what the entropy must be, we find
\be\label{S1}
S =
\frac{A}{4} + \frac{ \pi l^2 \ln  (r_+/l) }{4} \int \sinh \sigma d\sigma
\ee
if $d=4$ and
\be\label{S2}
S = \frac{A}{4} + \frac{\pi l^2 r_+^{d-4} }{4(d-4)}  \omega_{d-4} \int \sinh \sigma d\sigma
\ee
for $d > 4$.
The logarithm in the four-dimensional entropy here seems characteristic of a linear defect.  Indeed, the same type of term appears when this naive approach is used to calculate the entropy of spacetimes containing a Misner string, (e.g. the spherical NUT solution).  However, here we do not have a Misner string, so the origin of this term is puzzling.

It is interesting to note that, while both the mass and angular momentum in \eqref{thermos} have factors such as $\sinh^2 (\sigma/2)$ (and are therefore not simply proportional to the infinite volume of the hyperboloid sector), the combination ${\cal M}\equiv M - \Omega J$ can be written as
\[
{\cal M} = \frac{(d-2)m}{16} \omega_{d-4} \int \sinh \sigma d\sigma\,,
\]
which is proportional to the infinite volume.  This quantity corresponds precisely to what was treated as the mass of the hyperbolic Taub-NUT solution (treated as a non-rotating object) in \cite{Astefanesei:2004ji, Astefanesei:2004kn, Lee:2015wua} (cf. the discussion on pages 9 and 10 of \cite{Astefanesei:2004kn}) in the limit of $n = l/2$ and the rescalings  \eqref{rescaling}.  Obviously, the quantity ${\cal M}$,  obtained therein through the counter-term method,   corresponds to neither a mass nor an angular momentum.  However imposing
\be\label{Flawcallig}
\delta {\cal M}=T\delta S
\ee
 leads to the same entropies \eqref{S1} or \eqref{S2}, both proportional to the volume of the hyperboloid sector.

An alternative approach towards
analyzing the thermodynamics of this spacetime (that leads to a different entropy)  would be to enforce a relationship between the horizon radius and the AdS length (NUT parameter) that ensures the first law \eqref{Flawcallig}  holds \cite{Astefanesei:2004ji, Astefanesei:2004kn}.  This yields
the same periodicity as for spherical NUT charged spacetimes where $\beta = 8 \pi n$  ensures the absence of Misner strings \cite{Chamblin:1998pz, Johnson:2014xza}.  However, in the hyperbolic case, the validity of this constraint is dubious since there are no Misner strings present and hence there is no {\it a priori} reason to make this identification.  Nevertheless, it has been employed with some success in the thermodynamic description of  hyperbolic NUT spacetimes \cite{Astefanesei:2004ji, Astefanesei:2004kn, Lee:2015wua}.

Even if we include work terms due to angular momentum we do not obtain the usual area/entropy relationship in this alternative approach.
Specifically, suppose that the Euclidean time is periodic with period $\beta = 4\pi l / \alpha$ for some arbitrary (dimension dependent) constant $\alpha$ so that the temperature can then be written as
\be\label{extra}
T =  \frac{(d-5)l^2 + (d-1)r_+^2}{4 \pi r_+ l^2}  = \frac{\alpha}{4\pi l} \, .
\ee
This constraint forces a relationship between $r_+$ and $l$, which is explicitly
\be
l =  \frac{\alpha + \sqrt{\alpha^2 -(d-1)(d-5)}}{2|d-5|} r_+ \, .
\ee
Employing this constraint and enforcing the first law of thermodynamics \eqref{firstlaw}  for the mass and angular momentum given in \eqref{thermos}, we can calculate an entropy which satisfies \eqref{firstlaw}.  However, similar to the approach without  the extra condition \eqref{extra}, the resultant entropy evades the Bekenstein--Hawking area law.
The particular expressions are quite complicated and not especially illuminating.

To summarize, we have considered the thermodynamics of these rotating hyperboloid membranes and found that the expressions for mass, angular momentum and horizon area are divergent.  Furthermore, this is a problem that cannot be cured via compactification of the $(\sigma, \phi)$ sector due to the presence of rotation \cite{Klemm:1998kd}.  Nevertheless, we have found that the thermodynamic expressions satisfy the first law of thermodynamics  with the entropy given by the standard Bekenstein--Hawking result \emph{plus} an additional correction whose origin we have not understood.  Although a consistent thermodynamics can alternatively be obtained by enforcing a particular periodicity in Euclidean time  (analogous to what is done for the spherical NUT solution to eliminate Misner strings), this approach is both poorly motivated in the hyperbolic case
 and does not properly treat the rotational contribution to the first law \cite{Astefanesei:2004ji, Astefanesei:2004kn,Lee:2015wua}; indeed even if we include this contribution the usual entropy/area relation does not hold. Hence this approach does not provide any deeper understanding of the thermodynamics of these systems.

 Thermodynamic considerations seem even more complicated for generic hyperbolic Taub-NUT-AdS spacetimes, of which  \eqref{onespinlimit} is a special limit. Such solutions rotate non-uniformly, and  transformation to a non-rotating frame at infinity via a linear transformation of the Killing co-ordinates
 is not possible.  Hence transformation to a non-rotating frame (if it exists) will be  highly nontrivial
 and the formulation of the first law is at the least highly problematic \cite{Bardoux:2013swa}  (see, however, the approach in \cite{Astefanesei:2004ji, Astefanesei:2004kn, Lee:2015wua}). Similar issues also arise for `rotating topological  {black holes', see App.~A.} It is obvious that the thermodynamics of all these examples is burdened by similar issues and  warrants further investigation.

\section{A more general class of solutions}

In what follows we shall generalize the solution of the previous section,
applying the hyperboloid limit to the Kerr-AdS solution with multiple rotations, as well as combine it with the super-entropic limit, obtaining in particular solutions with all rotation parameters $a_i = l$.

\subsection{Kerr-AdS black holes}

We start from the general Kerr-AdS black hole \cite{GibbonsEtal:2004, GibbonsEtal:2005},
\begin{equation}
	\label{metric}
	ds^2
	=
	d\gamma^2+\frac{2m}{U} \omega^2+\frac{U dr^2}{F-2m}+d\Omega^2\,,
\end{equation}
where
\begin{align}
	d\gamma^2
	&=
	-\frac{W\rho^2}{l^2}dt ^2
	+ \sum_{i=1}^{N} \frac{r^2+a_i^2}{\Xi _i} \mu_i ^2 d\phi _i^2
	\,,\nonumber\\
	d\Omega^2
	&=
	\sum_{i=1}^{N+\varepsilon}\frac{r^2+a_i ^2}{\Xi _i} d\mu _i ^2
	-\frac{1}{W\rho^2}\Bigl(\sum_{i=1}^{N+\varepsilon}\frac{r^2+a_i ^2}{\Xi _i} \mu_i d\mu_i\Bigr)^2
	\,, \nonumber \\
	\omega
	&=
	W dt -\sum_{i=1}^{N} \frac{a_i \mu_i ^2 d\phi _i}{\Xi _i}
	\,,
\end{align}
and, as per usual $\rho^2=r^2+l^2$, while
\begin{align}
	\label{metricfunctions}
	W &= \sum_{i=1}^{N+\varepsilon}\frac{\mu _i^2}{\Xi _i}
	\,,\quad
	U=r^\varepsilon \sum_{i=1}^{N+\varepsilon} \frac{\mu _i^2}{r^2+a_i^2} \prod _j ^N (r^2+a_j^2)
	\,,\nonumber\\
	F&=\frac{r^{\varepsilon -2}\rho^2}{l^2}\prod_{i=1}^N (r^2+a_i^2)
	\,,\quad
	\Xi_i=1-\frac{a_i^2}{l^2}\,.\qquad
\end{align}
To treat even ($\varepsilon=1)$  and odd ($\varepsilon=0)$ spacetime dimensionality $d$ simultaneously, we have parametrized
\begin{equation}
d=2N + 1 + \varepsilon\,
\end{equation}
and in even dimensions set for convenience $a_{N+1}=0$.
The  ``direction cosines" $\mu_i$ are not independent, but obey the following constraint:
\begin{equation}
	\label{constraint}
	\sum_{i=1}^{N+\varepsilon}\mu_i^2=1\,
\end{equation}
in addition to $0 \le \mu_i \le 1$ for $1 \le i \le N$ and $-1 \le \mu_{N+1} \le 1$ (in even dimensions).

\subsection{Multi-hyperboloid membranes}

We can perform the hyperboloid membrane limit for $n$ rotation parameters by noting that the appropriate generalization of the substitution \eqref{eq:hyperlim_subs} is
\begin{align}
	\mu_k
	&=
	\sqrt{\Xi_k} \nu_k
	\label{hypersuper_subs}
\end{align}
for $k = 1, \ldots, n \le N+\varepsilon-1$.  Notice that we cannot have $k$ ranging all the way to $N+\varepsilon$ since then it would be impossible to satisfy eq. \eqref{constraint}.  In other words, there are $N+\varepsilon-1$ polar coordinates and we may perform the hyperboloid membrane limit for any number of these.

Upon making this substitution and taking the $a_k \to l$ limit, we find that
\begin{equation}
	\label{multiHBmetric}
	ds^2
	\to
	d\gamma_n^2+\frac{2m}{U_n} \omega_n^2+\frac{U_n dr^2}{F_n-2m}+d\Omega_n^2\,,
\end{equation}
with
\begin{align}
	d\gamma_n^2
	&=
	-\frac{W_n \rho^2}{l^2}dt ^2
	+ \rho^2 \sum_{k=1}^{n} \nu_k ^2 d\phi_k^2
	+ \sum_{i=n+1}^{N} \frac{r^2+a_i^2}{\Xi _i} \mu_i ^2 d\phi _i^2
	\,,\nonumber\\
	d\Omega_n^2
	&=
	\rho^2 \sum_{k=1}^{n} d\nu _k ^2
	+ \sum_{i=n+1}^{N+\varepsilon} \frac{r^2+a_i^2}{\Xi_i} d\mu_i^2
	\nn\\
	&-\frac{1}{W\rho^2}
	\Bigl(
		\rho^2 \sum_{k=1}^{n} \nu_i d\nu_i
		+ \sum_{i=n+1}^{N+\varepsilon}\frac{r^2+a_i ^2}{\Xi _i} \mu_i d\mu_i
	\Bigr)^2
	\,, \nonumber \\
	\omega_n
	&=
	W_n dt
	- l \sum_{k=1}^{n} \nu_k^2 d\phi_k
	-\sum_{i=n+1}^{N} \frac{a_i \mu_i ^2 d\phi _i}{\Xi _i}
	\,, \nonumber \\
	W_n
	&=
	\sum_{k=1}^{n} \nu_k^2
	+ \sum_{i=n+1}^{N+\varepsilon}\frac{\mu _i^2}{\Xi _i}
	\,, \nn \\
	F_n
	&=
	\frac{r^{\varepsilon -2}\rho^{2n+2}}{l^2} \prod_{i=n+1}^N (r^2+a_i^2)	
	\,,\nonumber\\
	U_n
	&=
	r^\varepsilon
	\sum_{i=n+1}^{N+\varepsilon} \frac{\mu _i^2}{r^2+a_i^2}
	\rho^{2n} \prod _{j=n+1}^N (r^2+a_j^2)
	\,,
\end{align}
and the constraint now reads
\begin{equation}\label{constraint}
	\sum_{i=n+1}^{N+\varepsilon}\mu_i^2=1\,.
\end{equation}
and the $\nu_k$ are now unbounded.

The class of metrics \eqref{multiHBmetric} describe multiply-rotating multi-hyperboloid membranes.
One does not have to take all the limits at once to arrive at  \eqref{multiHBmetric};
the same result can be obtained by taking the limits one at a time.

\subsection{Super-entropic hyperboloid membranes}

We now examine the superentropic limit of  \eqref{multiHBmetric} in the direction $\phi_j$ with $n < j \le N$.
As described in \cite{Hennigar:2014cfa, Hennigar:2015cja}   the superentropic limit is performed by transforming to a frame which rotates in the $j$ direction, rescaling $\varphi_j = \phi_j^R / \Xi_j$ and taking the limit $a_j \to l$.
Following the calculations in \cite{Hennigar:2015cja}, we find that the result is very similar,
\begin{equation}
	\label{hypersupermetric}
	ds^2
	\to
	d\gamma_{n,s}^2+\frac{2m}{U_{n,s}} \omega_{n,s}^2+\frac{U_{n,s} dr^2}{F_{n,s}-2m}+d\Omega_{n,s}^2\,,
\end{equation}
where we have defined
\begin{align}
	d\gamma_{n,s}^2
	&=
	-\big( (W_{n,s}+\mu_j^2) \rho^2 + \mu_j^2 l^2 \big) \frac{dt ^2}{l^2}
	+ 2 \rho^2 \frac{\mu_j^2}{l} dt d\varphi_j
	\nn\\
	&+ \rho^2 \sum_{k=1}^{n} \nu_k ^2 d\phi_k^2
	+ \sum_{\substack{i=n+1 \\ i \ne j}}^{N} \frac{r^2+a_i^2}{\Xi _i} \mu_i ^2 d\phi _i^2
	\,,\nonumber\\
	d\Omega_{n,s}^2
	&=
	\rho^2 \sum_{k=1}^{n} d\nu _k ^2
	+ \sum_{\substack{ i= n+1 \\ i \ne j}}^{N+\varepsilon} \frac{r^2+a_i^2}{\Xi_i} d\mu_i^2
	\nn\\
	&-2\frac{ d\mu_j}{\mu_j}
	\Bigl(
		\rho^2 \sum_{k=1}^{n} \nu_k d\nu_k
		+ \sum_{\substack{i = n+ 1 \\ i\neq j}}^{N+\varepsilon}\frac{r^2+  a^2_i }{\Xi _i} \mu_i d\mu_i
	\Bigr)
	\nn\\
	&+ \frac{d\mu _j ^2}{\mu _j ^2}\Bigl( \rho^2 W_{n,s} +l^2 \mu _j ^2 \Bigr)
	\,, \nonumber \\
	\omega_{n,s}
	&=
	(W_{n,s}+\mu^2_j) dt
	- l \mu_j^2 d\varphi_{j}
	\nn \\
	&- l \sum_{k=1}^{n} \nu_k^2 d\phi_k
	-\sum_{\substack{i=n+1 \\ i\neq j}}^{N} \frac{a_i \mu_i ^2 d\phi_i}{\Xi _i}
	\,, \nonumber \\
	W_{n,s}
	&=
	\sum_{k=1}^{n} \nu_k^2
	+ \sum_{\substack{i=n+1 \\ i \ne j}}^{N+\varepsilon}\frac{\mu _i^2}{\Xi _i}
	\,,
	\nn\\
	U_{n,s}
	&=
	r^\varepsilon
	\bigg(
		\sum_{i \ne j}^{N+\varepsilon} \frac{\mu _i^2 \rho^2}{r^2+a_i^2}
		+ \mu_j^2
	\bigg)
	\rho^{2n} \prod _{i_2 \ne j}^N (r^2+a_{i_2}^2)
	\,,\nonumber\\
	F_{n,s}
	&=
	\frac{r^{\varepsilon -2} \rho^{2n+4}}{l^2} \prod_{i=n+1}^N (r^2+a_i^2)
	\,,
\end{align}
and the constraint still reads
\begin{equation}\label{eqn:constraintHyperSuper}
	\sum_{i=n+1}^{N+\varepsilon}\mu_i^2
	=1\,.
\end{equation}
This is the same as what one would obtain by applying the $n$-fold hyperboloid membrane limit to the general super-entropic black hole found in \cite{Hennigar:2015cja}.

\subsection{Examples \& basic properties}

A few comments are in order with regard to the metrics~\eqref{multiHBmetric} and \eqref{hypersupermetric}.  In previous work \cite{Hennigar:2014cfa, Hennigar:2015cja}  which focused on the super-entropic limit, it was found impossible to perform more than a single super-entropic limit.  Furthermore, after taking the super-entropic limit, it was always the case that punctures were introduced into black hole horizon.  We have  seen  already that the first of these points does not apply to the hyperboloid membrane limit---it is possible to perform the hyperboloid membrane limit as many times as there are polar angles.  In particular this means, in even dimensions (where the number of polar angles equals the number of rotation parameters), one can obtain a solution where all $a_i=l$ by performing the hyperboloid membrane limit for all polar angles.

When considering the combination of some number of hyperboloid membrane limits and the super-entropic limit, we can still observe the characteristic punctures since the coefficient of $d \mu_j^2$ in eq. \eqref{hypersupermetric} diverges as $\mu_j \to 0$.  However, a striking feature of odd dimensional solutions is the possibility of having no punctures.  In odd dimensions, there are $N$ rotational parameters and $N-1$ polar angles. If in \eqref{hypersupermetric} we choose $n=N-1$, then the constraint \eqref{eqn:constraintHyperSuper} reduces to $\mu_j^2 = \mu_{N}^2 = 1$, and we see that the terms responsible for the punctures drop out of the solution.  In other words, for an odd dimensional solution where the maximum number of hyperboloid membrane limits have been performed, taking the super-entropic limit for the remaining rotational parameter will not introduce punctures.  We discuss this feature explicitly for the $d=7$ solution in the next subsection.

\begin{figure}[htp]
\includegraphics[scale=0.5]{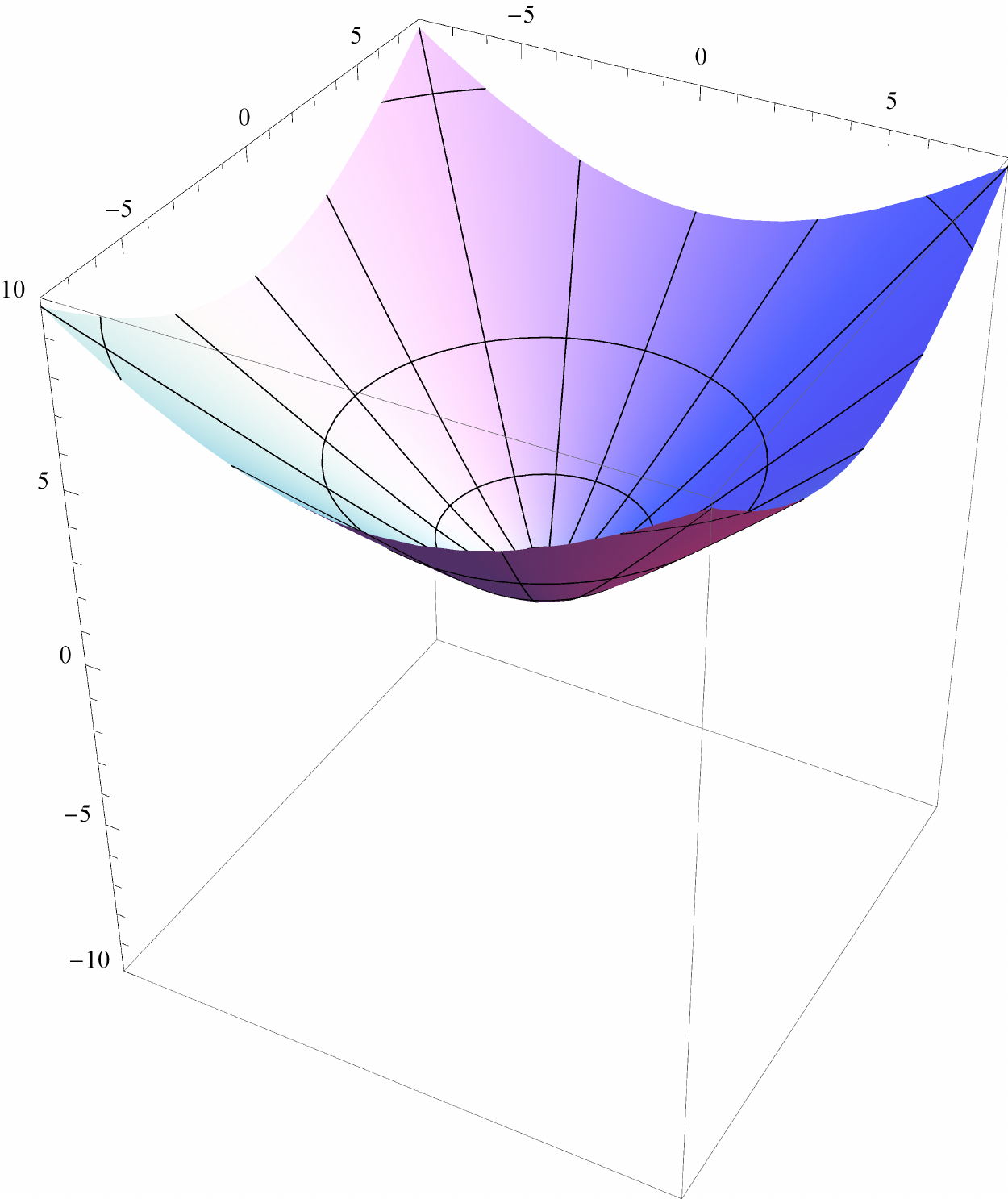}
\caption{
Embedding of the $(\sigma, \phi)$ sector of metric \eqref{onespinlimit} for $r_+=\sqrt{10}, l=1$.  Here we see that the horizon is a hyperbola.}
\end{figure}

We now write explicitly a few examples of these metrics in more familiar coordinates, for both the even and odd dimensions. In particular, we focus on cases where all rotational parameters have been set equal to the AdS length.  Since the $5$-dimensional solution of this form was presented in \cite{Hennigar:2015cja} we turn first to $6$ dimensions.

\subsubsection{Solutions in $6d$}
In $6$ dimensions there are two polar coordinates and two rotational parameters.  There are two ways in which both rotational parameters can be taken to $l$: (i) two hyperboloid membrane limits, or (ii) one hyperboloid membrane limit combined with the super-entropic limit.

(i) In the case of two hyperboloid membrane limits, the metric is given by eq. \eqref{multiHBmetric} and, in more familiar coordinates, can be written as
\ba
&&ds^2 = -\rho^2\left(1+ \sinh^2\!\sigma +  \sinh^2\!\theta \right) \frac{dt^2}{l^2} + \frac{\rho^4 l^2 dr^2}{\rho^6-2mrl^2}
\nn\\
&&+ \rho^2\left(\sinh^2\!\sigma d\phi^2 +\sinh^2\!\theta d\psi^2 \right)
+ \frac{2mr}{\rho^4}\omega^2   + d\Omega^2\,,
\ea
with
\ba
\omega &=& \left[1+ \sinh^2\!\sigma + \sinh^2\!\theta \right] dt
\nn\\
&&-l\left(\sinh^2\!\sigma d\phi + \sinh^2\!\theta d\psi \right) \, ,
\nn\\
d\Omega^2 &=& \rho^2\left(\cosh^2\!\sigma d\sigma^2 + \cosh^2\!\theta d\theta^2 \right)
\nn\\
&&- \rho^2 \frac{\left[\cosh\sigma \sinh\sigma d\sigma + \cosh\theta\sinh\theta d\theta \right]^2}{1+ \sinh^2\!\sigma + \sinh^2\!\theta} \, .
\ea
 The solution admits a horizon when $\rho^6 - 2mrl^2=0$, in which case it describes an asymptotically AdS black membrane.

Rescaling the above metric by $(l/r)^2$ and taking $r \to \infty$ we find that the boundary metric for this spacetime is,
\ba
ds^2_{\rm bdry} &=& -\left(1+ \sinh^2\!\sigma +  \sinh^2\!\theta \right) dt^2
 \\
&&+ l^2\left(\sinh^2\!\sigma d\phi^2 +\sinh^2\!\theta d\psi^2 \right) + l^2 d\Omega^2_{\rm bdry}\,, \nn
\ea
where
\ba
d\Omega^2_{\rm bdry} &=& \left(\cosh^2\!\sigma d\sigma^2 + \cosh^2\!\theta d\theta^2 \right)
\nn\\
&&- \frac{\left[\cosh\sigma \sinh\sigma d\sigma + \cosh\theta\sinh\theta d\theta \right]^2}{1+ \sinh^2\!\sigma + \sinh^2\!\theta} \, .
\ea
 It is straightforward to check that this boundary metric is in fact nothing more than AdS$_5$ written in unusual coordinates.
 
The induced metric on the horizon, which, after the transformation $(R_1,R_2, \Phi_1, \Phi_2) = (\sinh\sigma, \sinh\theta, \phi,\psi)$, reads
\ba
ds_{\text{hor}}^2 &=& \rho_+^2 \bigg[dR_1^2 + dR_2^2 + R_1^2d\Phi^2_1 + R_2^2 d\Phi_2^2
\nn\\
&+& (R_1^2d\Phi_1 + R_2^2d\Phi_2)^2 - \frac{(R_1dR_1 + R_2dR_2)^2}{1+ R_1^2+R_2^2} \bigg] \,.
\nn\\
\ea
This is a space whose Ricci scalar  (${\cal R} = - 24/ \rho^2_+$)  and  Kretschmann scalar $(K = 192/\rho_+^4$) are both constant.

(ii) The solution obtained by taking one hyperboloid and one super-entropic limit is given by eq. \eqref{hypersupermetric} and reads
\ba\label{hypersuper6d}
ds^2 &=& -\left(\rho^2\cosh^2\!\sigma  + l^2 \sin^2\!\theta \right)\frac{dt^2}{l^2} + \frac{2\rho^2 \sin^2\!\theta}{l} dt d\phi
\nn\\
&&+ \rho^2\sinh^2\!\sigma d\psi^2  - \frac{2\rho^2\cosh\sigma\sinh\sigma\cos\theta}{\sin\theta} d\sigma d\theta
\nn\\
&&+ \rho^2\cosh^2\!\sigma d\sigma^2  + \frac{\rho^2\cosh^2\!\sigma \cos^2\!\theta+r^2\sin^2\!\theta}{\sin^2\!\theta} d\theta^2
\nn\\
&&+ \frac{2mr\left[\cosh^2\!\sigma dt -l\sin^2\!\theta d\phi -l\sinh^2\!\sigma d\psi \right]^2}{\rho^2(\rho^2-l^2\sin^2\!\theta)}
\nn\\&&+ \frac{l^2\rho^2(\rho^2-l^2\sin^2\!\theta)}{\rho^6 -2mrl^2}dr^2\,,
\ea
which is again asymptotically  AdS.

\begin{figure}[htp]
\includegraphics[scale=0.5]{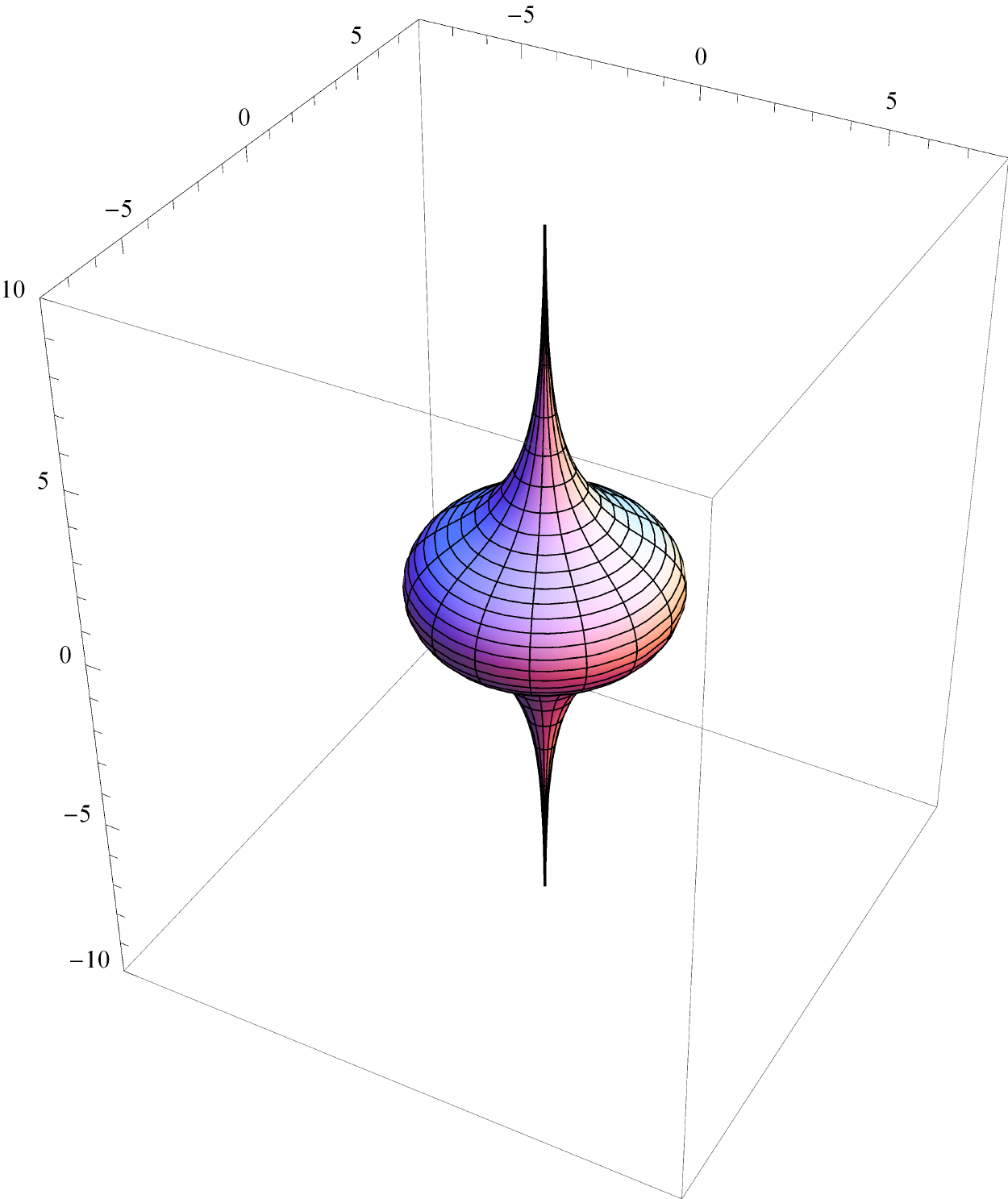}
\caption{
	Embedding of the $(\theta, \phi)$ sector of metric \eqref{hypersuper6d} for $r_+=\sqrt{10}, l=1, \sigma=0$ and $\phi \sim \phi + 2\pi$.  For this slice we see the characteristic shape of the super-entropic black holes.  Increasing the value of $\sigma$ effectively stretches the plot vertically.
}
\label{punctures}
\end{figure}

We see in this case the characteristic punctures (referring to the $d\theta$ terms in the metric) associated with the super-entropic limit.  These are present because here we have not taken the maximum possible number of hyperboloid membrane limits, but rather chose to set one rotational parameter to $l$ using the super-entropic limit.  To visualize the punctures, we can embed the $(\theta, \phi)$ sector of the horizon in $\mathbb{E}^3$, with the result shown in Figure~\ref{punctures}. On the horizon, neither the Ricci nor  Kretschmann scalars are constant---both are functions of $\theta$.

To consider the boundary geometry of this solution, we again rescale by $(l/r)^2$ and take $r \to \infty$ obtaining
\ba
ds^2_{\rm bdry} &=& -\cosh^2 \sigma dt^2 + 2 l \sin^2\!\theta dt d\phi + l^2 \sinh^2 \sigma d\psi^2
\nn\\
&& - \frac{2 l^2 \cosh \sigma \sinh \sigma \cos \theta}{\sin \theta} d\sigma d\theta + l^2 \cosh^2 \sigma d\sigma^2
\nn\\
&& + \frac{l^2 (\cosh^2\!\sigma \cos^2\!\theta + \sin^2 \theta)}{\sin^2\!\theta} d\theta ^2
\ea
This boundary metric has neither Ricci nor  Kretschmann scalars constant.  In fact, both depend on the angular coordinate $\theta$  as follows
\ba
{\cal R} &=& -\frac{4}{l^2} \left(7 \cos^2\!\theta - 2 \right)\,,
\nn\\
K &=& \frac{8}{l^4}\left(11 \cos^4\!\theta - 8\cos^2\!\theta + 2 \right) 
\ea
and it is straightforward to show that the Weyl tensor vanishes, indicating the boundary is conformally flat. We were unable to determine the conformal factor.

\subsubsection{Solutions in $7d$}

Turning now to $d=7$ dimensions, for which there are up to three rotation parameters and two polar coordinates, plenty of possibilities emerge. However, to obtain all $a_i=l$ requires the unique combination of  two hyperboloid membrane limits with a super-entropic limit.  The resulting metric is
\ba\label{hypersuper7d}
ds^2 &=& -\left[ \rho^2 \left(1+\sinh^2\!\theta+\sinh^2\!\sigma \right) +l^2 \right]\frac{dt^2}{l^2} + \frac{2\rho^2 dtd\xi}{l}
\nn\\
&&+\rho^2 \left(\sinh^2\!\sigma d\phi^2 + \sinh^2\!\theta d\psi^2 \right) + \frac{\rho^4 r^2 l^2 dr^2}{\rho^8 - 2mr^2l^2}
\nn\\
&&+ \frac{2m}{\rho^4} \omega^2 + \rho^2 \left(\cosh^2\!\sigma d\sigma^2 + \cosh^2\!\theta d\theta^2 \right)
\ea
with
\ba
\omega &=& \left[1+\sinh^2\!\sigma + \sinh^2\!\theta \right] dt - ld\xi
\nn\\
&& -l\left(\sinh^2\!\sigma d\phi + \sinh^2\!\theta d\psi \right)
\ea
where  $\xi$ is the azimuthal coordinate associated with the super-entropic limit; it may be considered compact, with $\xi \sim \xi + \mu$.  The spacetime is  asymptotically AdS.

The metric on the conformal boundary is
\ba
ds^2_{\rm bdry} &=& -\left( 1 +\sinh^2\!\theta+\sinh^2\!\sigma \right) dt^2 + 2 l dt d\xi
\nn\\
&&+ l^2\left(\sinh^2\!\sigma d\phi^2 + \sinh^2\!\theta d\psi^2 \right)
\nn\\
&& + l^2\left(\cosh^2\!\sigma d\sigma^2 + \cosh^2\!\theta d\theta^2 \right)
\ea
and has vanishing Ricci and  Kretschmann scalars.  Again, the  Weyl tensor vanishes, and the boundary is conformally flat.

Since the maximum number of hyperboloid membrane limits have been taken, we expect that the super-entropic limit has not introduced any punctures.  To demonstrate this explicitly, consider the induced metric on the horizon.   Using the transformation $(R_1,R_2,\Phi_1,\Phi_2)=(\sinh\sigma,\sinh\theta,\phi,\psi)$ we can write the induced metric on the horizon as
\ba
ds_{\textrm{hor}}^2 &=& \rho_+^2 \left[dR_1^2 + R^2_1 d\Phi^2_1 + dR_2^2 + R^2_2 d\Phi^2_2  \right. \nn\\
&&\left. + \frac{\rho_+^2}{r_+^2}\left(d\xi + R^2_1 d\Phi_1 + R^2_2 d\Phi_2
\right)^2  \right]
\ea
which is the fibration of a circle over a four-dimensional space of non-constant curvature and constant Kretschmann scalar.   We find the Ricci scalar is given by ${\cal R} = -4\rho_+^2/r^2_+$, while the Kretschmann scalar is $K = 136\rho_+^4/r^4_+$.
However the space, though nonsingular, is not a space of constant curvature.

\section{Conclusions}

 In this paper we have studied the hyperboloid membrane limit both as known in the literature
\cite{Caldarelli:2008pz, Caldarelli:2012cm} and generalized to a multi-hyperboloid limit  and/or combined with the super-entropic limit, obtaining in particular rotating black holes for which all rotational parameters are  equal to the AdS radius.

We have dedicated much effort to the analysis of the thermodynamics of the singly-spinning solutions.  Such solutions correspond to a particular limit of hyperbolic Taub-NUT-AdS spacetimes, where the NUT parameter attains the value $n=l/2$.  The thermodynamics of these solutions have been misrepresented in the literature, since their rotation has been neglected.  Including the appropriate angular momentum terms, we find that the first law is only satisfied if the entropy takes a form different from the standard area law.
This happens despite the fact that a Misner string is not present; the additional assumption on periodicity of the Euclidean time \'ala \eqref{extra} does not `cure' this anomaly.
Similar issues arise in thermodynamics of hyperbolic Taub-NUT-AdS spacetimes or rotating topological black holes
and deserve further attention.

The new multiply-rotating multi-hyperboloid membranes we have constructed are solutions to vacuum Einstein equations with negative cosmological constant.   The multi-hyperboloid membranes differ from the previously studied super-entropic black holes in that it is possible to take an ultraspinning limit in all possible directions, whereas the super-entropic limit can only be applied for a single rotational parameter.
In some cases, applying the hyperboloid membrane limit to the multiply-rotating super-entropic black holes of \cite{Gnecchi:2013mja,Klemm:2014rda,Hennigar:2014cfa,Hennigar:2015cja} removes their characteristic puncture.

The solutions we have obtained provide a tool for the study of fluids on generalized rotating Einstein universes.  
Possible applications of these metrics to holography and a deeper understanding of their thermodynamics are left for future studies.

\section*{Acknowledgments}
We would like to thank R.~Myers for a discussion.
This research was supported in part by Perimeter Institute for Theoretical Physics and by the Natural Sciences and Engineering Research Council of Canada. Research at Perimeter Institute is supported by the Government of Canada through Industry Canada and by the Province of Ontario through the Ministry of Research and Innovation.

\appendix

\section{Rotating topological black holes}
In this appendix we  review the singly-spinning $d$-dimensional `rotating topological black holes' constructed by Klemm et. al.  \cite{Klemm:1997ea, Klemm:1998kd}. The solutions can be obtained by the following analytic continuation of Kerr-AdS spacetimes \eqref{KerrAdSsingle}:
\be
t\to it\,, \quad r\to ir\,,\quad \theta\to i\theta\,,\quad  a\to ia\,,\quad m\to -i^{d-3}m\,.
\ee
The resulting metric reads
\ba
ds^2 &=& -\frac{\tilde{\Delta}_a}{\tilde{\rho}^2_a}\left[dt + \frac{a}{\tilde{\Xi}_a} \sinh^2\!\theta d\phi \right]^2 \!\!\!+\frac{\tilde{\rho}^2_a}{\tilde{\Delta}_a}dr^2 + \frac{\tilde{\rho}^2_a}{\tilde{\Sigma}_a}d\theta^2\qquad  \label{rottop}\\
&+& \frac{\tilde{\Sigma}_a \sinh^2\!\theta }{\tilde{\rho}_a^2} \left[a dt - \frac{r^2+a^2}{\tilde{\Xi}_a} d\phi \right]^2\!\!\! +r^2 \cosh^2\!\theta d\hat{\Sigma}_{d-4}^2\,, \qquad\nn
\ea
where
\ba
\tilde{\Delta}_a &=& (r^2+a^2)(k+\frac{r^2}{l^2}) - 2mr^{5-d}, \quad \tilde{\Sigma}_a = 1+\frac{a^2}{l^2}\cosh^2\!\theta\,,\nn\\
\tilde{\Xi}_a &=& 1+\frac{a^2}{l^2}, \quad \tilde{\rho}^2_a = r^2+a^2\cosh^2\!\theta\,.
\ea
Here $k=-1$ and a further analytic continuation has been carried out so that
$d\Omega_{d-4}^2 \to d\hat{\Sigma}_{d-4}^2$, where $d\hat{\Sigma}_{d-4}^2$ is the metric of a compact hyperbolic space \cite{Dehghani:2002wp}. The constant $(t,r)$ hypersurfaces of the resulting solution (\ref{rottop}) have topology $\mathbb{H}^2 \times \mathbb{H}^{d-4}$ and are in many
ways distinct from the hyperboloid membranes studied in Sec.~\ref{Sec2}.

For example, in $d=4$, the metric (\ref{onespinlimit})
will describe a naked singularity for sufficiently small $m$, a black hole with an outer and an inner horizon for sufficiently large $m$, and an extremal black hole for the critical value $m=8\sqrt{3}l/9$; however a $d=4$ rotating topological black hole \cite{Klemm:1997ea} (in the $a\to l$ limit) will have a double-horizon causal structure for any value of $m>0$.

 It was demonstrated in \cite{Klemm:1998kd} that in the presence of rotation, the $(\theta,\phi)$ sector cannot be compactified (without introducing a discontinuity to the metric) and the solutions in fact represent {\em rotating black branes} rather than topological black holes.
The thermodynamics of these objects has been studied in \cite{Klemm:1997ea, Dehghani:2002wp} and seems to suffer from {serious problems, some related to those studied in Sec.~II.}

 In particular, concentrating on four dimensions, the transformation $\phi\to \phi+\frac{a}{l^2}t$ brings the metric to a non-rotating frame.
Using the conformal method we then find the {following expressions for divergent} thermodynamic quantities:
\ba
M&=&\frac{m}{4\Xi^2}\int\Bigl[2\Xi+3(a^2/l^2)\sinh^2\!\theta\Bigr]\sinh\theta d\theta\,,\nonumber\\
J&=&-\frac{3}{4}\frac{ma}{\Xi^2}\int \sinh^3\!\theta d\theta\,,\quad
\Omega=\frac{a(l^2-r_+^2)}{(r_+^2+a^2)l^2}\,,\nonumber\\
T&=&\frac{3r_+^4+(a^2-l^2)r_+^2+a^2l^2}{4\pi l^2r_+(r_+^2+a^2)}\,,\nonumber\\
A&=&2\pi\frac{r_+^2+a^2}{\Xi}\int \sinh\theta d\theta\,.
\ea
 However, the first law \eqref{firstlaw} is not satisfied for any $S$ as long as both $r_+$ and $a$ are independently varied.
This problem warrants further attention \cite{Klemm:1997ea}.

\section{Singly-spinning super-entropic black holes}
The $d$-dimensional singly spinning super-entropic black holes have been constructed in \cite{Hennigar:2014cfa}, extending the 4-dimensional work of \cite{Gnecchi:2013mja, Klemm:2014rda}. The metric reads
\ba\label{Singlyspinning}
ds^2&=&-\frac{\Delta}{\rho^2}(dt-{l}\sin^2\!\theta d\psi)^2+
\frac{\rho^2}{\Delta}dr^2+\frac{\rho^2}{\sin^2\!\theta}d\theta^2\quad \nonumber\\
&+&
\frac{\sin^4\!\theta}{\rho^2}[ldt-({r^2\!+\!l^2})d\psi]^2\!+\!r^2\cos^2\!\theta d\Omega_{d-4}^2\,,\qquad\ \
\ea
where
\ba
\Delta&=&\Bigl(l+\frac{r^2}{l}\Bigr)^2-2mr^{5-d}\,,\ \rho^2=r^2+l^2\cos^2\!\theta\,,\quad
\ea
and $d\Omega_{d}^2$ denotes the metric element on a $d$-dimensional sphere and coordinate $\psi$ is compactified according to $\psi \sim \psi + \mu$.

 Despite the fact that the horizon has a puncture and hence is non-compact, the thermodynamics is well defined for these black holes.
Namely, the following quantities satisfy the first law $\delta M=T\delta S+\Omega \delta J+V\delta P$:
\begin{eqnarray}\label{singlespin}
M&=&\frac{\omega_{d-2}}{8\pi} \left(d-2 \right)m  \,, \quad
J=\frac{2}{d-2}Ml\,,\quad \Omega=\frac{l}{r_+^2+l^2}\,,\nonumber\\
T&=&\frac{1}{4\pi r_+ l^2}\Bigr[ (d-5)l^2 + r^2(d-1)\Bigr]\,,\quad P =\frac{(d-1)(d-2)}{16\pi l^2}\,,\nonumber\\
S&=&\frac{\omega_{d-2}}{4}(l^2+r_+^2) r_+^{d-4}=\frac{A}{4}\,,\quad
V=\frac{r_+A}{d-1}\,,
\end{eqnarray}
where
\be
\omega_{d} = \frac{\mu \pi ^{\frac{d-1}{2}}}{\Gamma\left(\frac{d+1}{2}\right)}
\ee
is the volume of the $d$-dimensional  unit `sphere'. The horizon topology is that of a sphere with two punctures.


\providecommand{\href}[2]{#2}\begingroup\raggedright\endgroup

\end{document}